\documentclass[preprint2]{aastex}
\usepackage{float}
\usepackage{color}
\usepackage{url} 
\usepackage{hyperref} 
\usepackage{graphicx}

\title{The metallicity sensitivity of a surface brightness temperature scale}

\author{{Jeremy Mould$^{1}$}\\
\affil{$^1$Centre for Astrophysics \& Supercomputing, Swinburne University, PO Box 218, Hawthorn, Vic 3122, Australia}}

\hypersetup{colorlinks,citecolor=blue,linkcolor=blue,urlcolor=blue}

\begin{document}
\begin{abstract}
To obtain the accuracy now sought in the extragalactic distance scale through standard candles and rulers, calibration of stellar photometry must be improved. The sensitivity of the V--K color surface brightness relation is examined here by means of model atmosphere fluxes. It has previously been neglected, but is shown here to be a significant term in the error budget of a recent high precision distance of the Large Magellanic Cloud, an anchor in galaxy distances based on Cepheids.
\end{abstract}

\maketitle
\section{Introduction}
The goal of 1\% accuracy in the extragalactic distance scale places new demands on stellar astrophysics. ESA's $Gaia$ mission has opened up new possibilities in the classes of stars that can be standard candles (e.g. Mould, Clementini, Da Costa 2019). Pietrzynski et al (2019) employ the surface brightness colour relation to attain the required accuracy in the distance of the Large Magellanic Cloud (LMC). Both works require accurate calibration of photometry to temperatures and luminosities.

In this paper we explore the metallicity dependence of the surface brightness color relation using the surface fluxes of Kurucz model atmospheres.

\section{Surface brightness and effective temperature}
Eclipsing binaries allow accurate measurements of stellar radii (e.g. Elgueta et al 2018). Physically, we can understand this as combining the definition of effective temperature T$_e$
$$L = 4\pi \sigma R^2 T_e^4$$
where L and R are the stellar luminosity and radius respectively, with a geometric measurement of R. Dividing by stellar surface area, we obtain a flux
$$F =  \sigma \theta^2 T_e^4$$
where $\theta$ is the stellar angular radius and $\sigma$ is the Stefan-Boltzmann constant.
If the flux is measured photometrically, the effective temperature can be
estimated from V--K (e.g. di Benedetto et al 1998, 2005). One combines $\theta$ and R to obtain the distance.

In these terms we can express the error propagation: $\delta \theta/\theta ~=~ 2 \delta T_e/T_e$ 
If V--K = $f$(T$_e$, g, Z), then $$\delta (V-K) = \frac{\partial f}{\partial T_e} \delta T_e + \frac{\partial f}{\partial g} \delta g + \frac{\partial f}{\partial Z} \delta Z$$
It is possible to ignore the second term for the time being, supposing the ratio of stellar mass to R$^2$ to be perfectly determined by a spectroscopic eclipsing binary solution and investigate the third term using the predictions of Kurucz model atmospheres.
\begin{figure}[H]
\includegraphics[width=.35\textwidth, angle=-90]{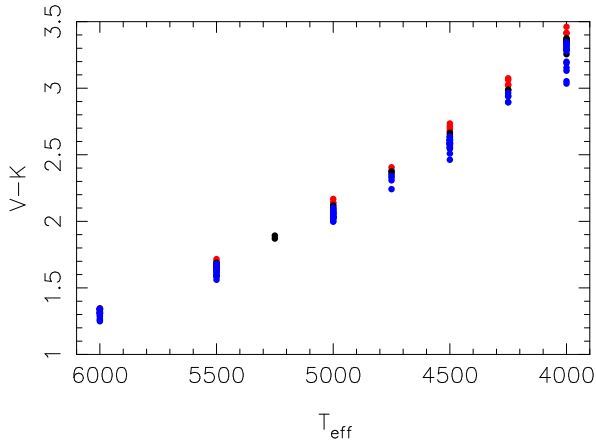}
\caption{V--K colour vs temperature. Red symbols are metal rich; blue: metal poor. Colors are calculated as filter integrated flux ratios and normalised to the Sun.}
\end{figure}
\section{Synthetic Color Temperature relation}
The relation between V--K and T$_e$ can be modeled using
Buser \& Kurucz (1992) fluxes and Bessell (2005), together with Bessell \& Brett
(1988), filter responses. 
The fluxes of these models are not well sampled in the near infrared
and were therefore interpolated in the K bandpass. Results are independent of whether linear or parabolic interpolation was used.
The dependence of color on metallicity at fixed gravity is illustrated in Figure~1. Similar results are obtained using model atmosphere fluxes with nearly two orders of magnitude more spectral resolution (Allard 2016). These models include TiO in the V bandpass and CO in the K bandpass, and so are superior in their predictions for K giants and M stars.
\begin{figure}[H]
\includegraphics[width=.35\textwidth, angle=-90]{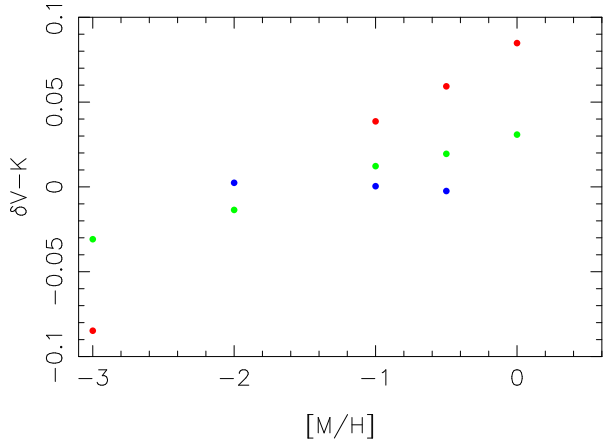}
\caption{Modeled V--K color dependence on metallicity for three stellar temperatures and fixed gravity. The metallicity [M/H] is the logarithmic Z value relative to the Sun. Red symbols are T$_e$ = 4500K. Green: 5000K; blue: 5500K.}
\end{figure}
Spherical models are also available to complement the standard plane parallel atmosphere models (e.g. SATLAS, Lester \& Neilson 2008). At one solar mass MARCS models\footnote{marcs.astro.uu.se} by Gustafsson et al (2008) are redder in V--K at 5000K and log g = 3 in the spherical case than the plane parallel case, but $\partial f/\partial\log Z$ is almost identical.
\section{Red clump stars}
Pietrzynski et al (2019) employ eclipsing binaries from the red clump in the LMC. Onozato et al (2019) find a mean value K $\approx$ 16.82 mag for the red clump
in star clusters, and if the LMC is 50 kpc distant, a bolometric correction BC$_K$ = 1.92 mag (Johnson 1966) gives log~L/L$_\odot~\approx$ 2.0. For solar mass stars the gravity is log~g = 2.2.
Figure 2 shows that d(V-K)/dlogZ $\approx$ 0.2/3 mag/dex.% approximately.

If d(V-K)/dlogT$_e$ = 2/0.176, then dlogT$_e$ / dlogZ = 0.1 $\times$ 0.176
This means that if an error $\delta$logZ = 0.3 is made, then $\delta$logT$_e$ =0.434 $\delta$T/T = 1.76 $\times$ 10$^{-3}$
and $\delta$T/T = 0.004, corresponding to a 0.8\% error in distance.
\vfill\break
\section{Giant Branch Stars}
If the stars lie on a giant branch, appropriate gravities for each temperature should be employed.
In solar logarithmic units [g] = [M] -- [L] + 4[T$_e$].
Modeling the giant branch as a one solar mass line commencing at the Sun and linear in log L, log T to L$_{tip}$, T$_{tip}$, we can write
[L] = [L$_{tip}$]/[T$_{tip}$] [T$_e$]
If L$_{tip}$ = 2000 L$_\odot$ 
 and T$_{tip}$ = 3500K, then [g] = 18.9 [T$_e$]
With these assumptions we derive Figure 3. 
\begin{figure}[h]
\includegraphics[width=.35\textwidth, angle=-90]{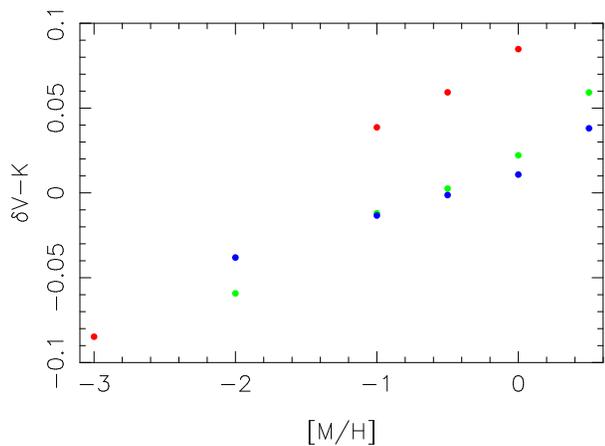}
\caption{Modeled V--K color dependence on metallicity for three stellar temperatures on the schematic red giant branch outlined in $\S$5. %The metallicity [M/H] is the logarithmic Z value relative to the Sun. R
As in the previous figure, red symbols are T$_e$ = 4500K. Green: 5000K; blue: 5500K.}
\end{figure}
\section{Main sequence stars}
The surface brightness V--K relation is also employed on the main sequence.
We show the modeled gravity dependence of V--K, the second term in the error propagation equation, in Figure 4.
\begin{figure}
\includegraphics[width=.35\textwidth, angle=-90]{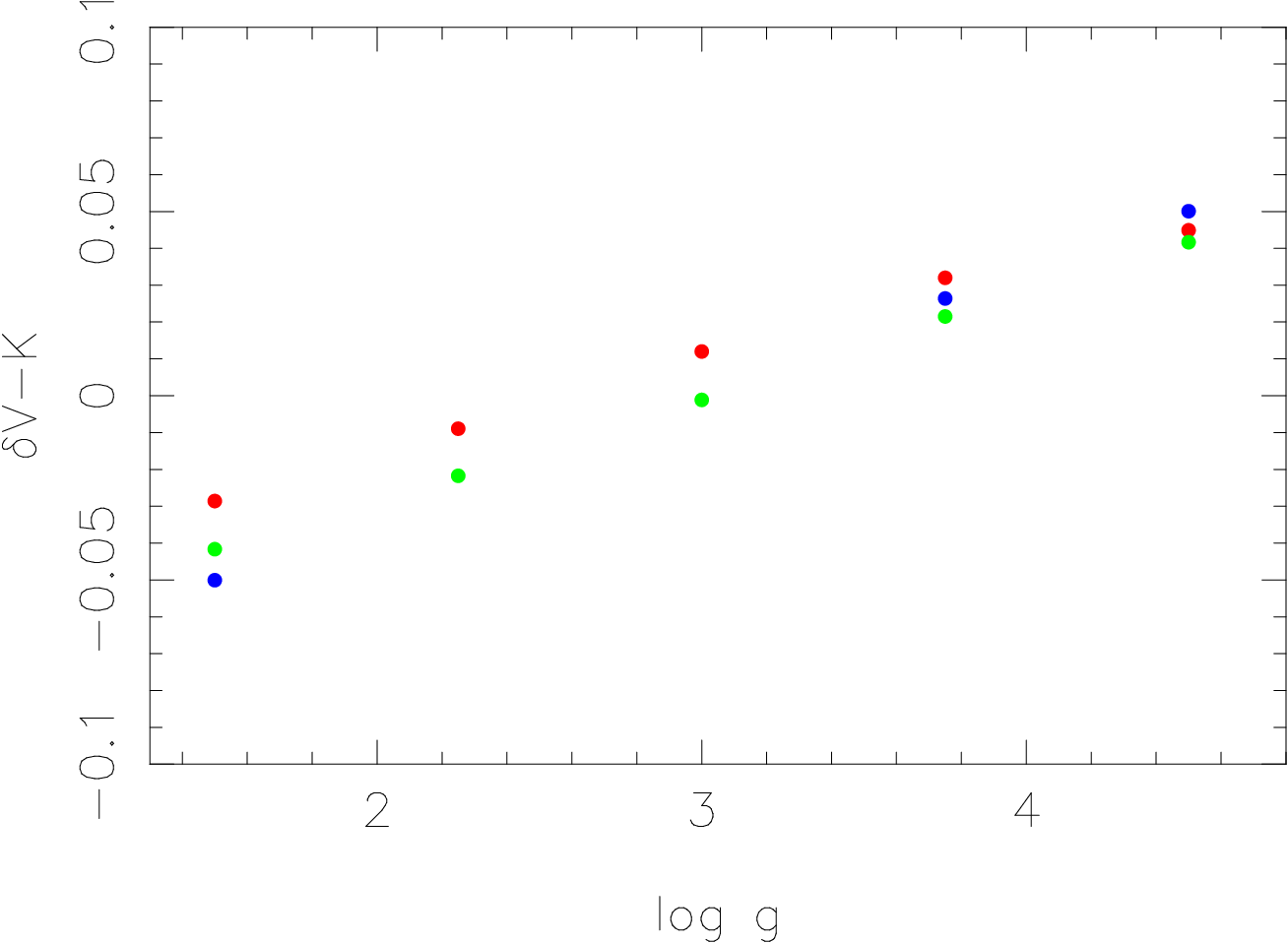}
\caption{Modeled V--K color dependence on gravity for three stellar temperatures at solar metallicity (above) and approximately one third solar metallicity (below).
As in the previous figures, red symbols are T$_e$ = 4500K. Green: 5000K; blue: 5500K.}
\includegraphics[width=.35\textwidth, angle=-90]{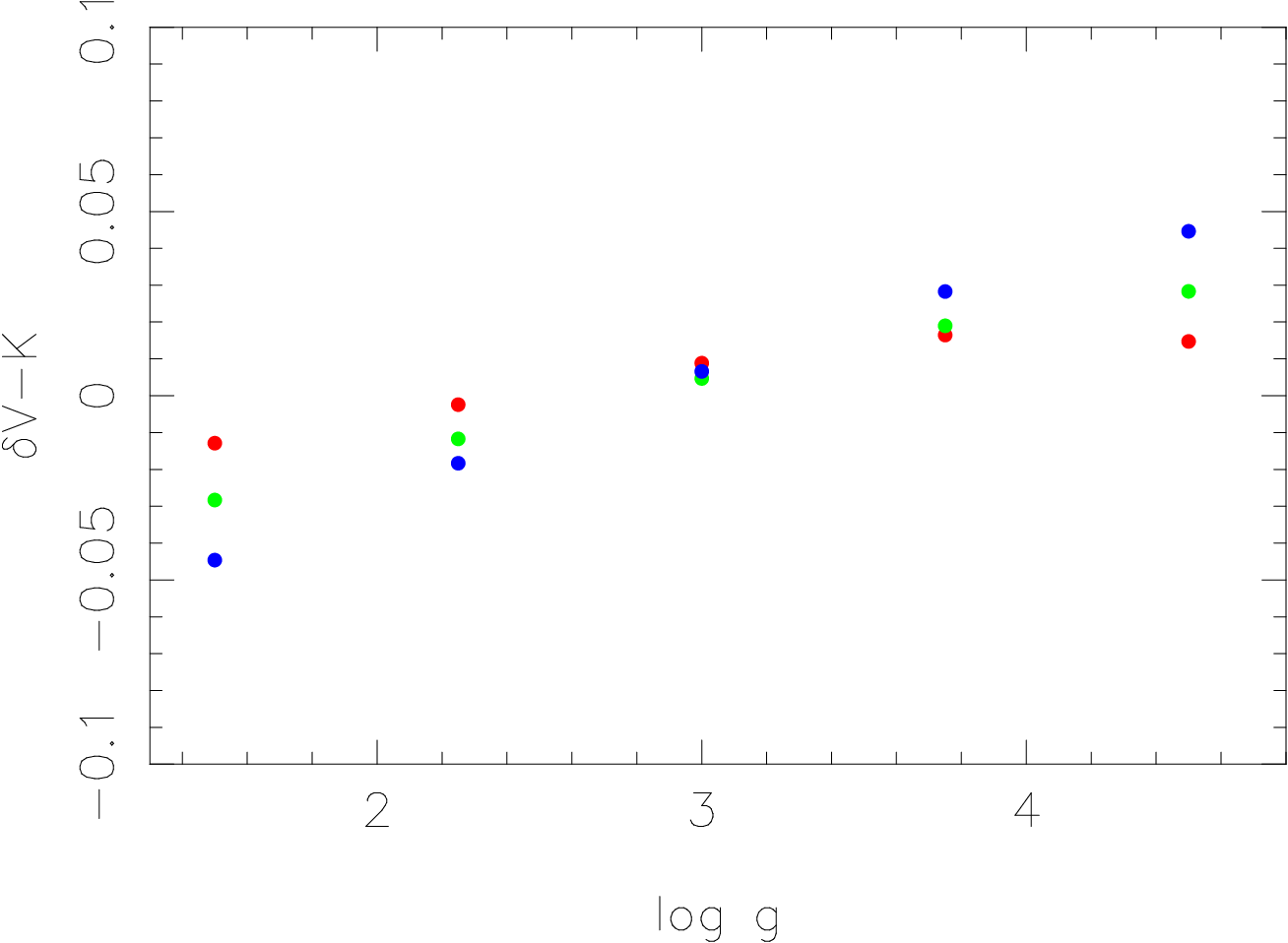}
\end{figure}

\section{Conclusion}
For precision of 1\% in distance the metallicities of eclipsing binaries should not be neglected, as the surface brightness color relation is affected by V--K metallicity dependence at this level. In the range they explore, Onozato et al (2019) find that population effects on JHK colors of LMC red clump stars are small. It may be possible to control the metallicity dependence noted here by small correction to the V--K colors in the V bandpass where blanketing by absorption lines is greater than in the infrared.

I would like to thank the referee for useful comments.

\section*{References}
\noindent Allard, F. 2016, SF2A, {\it Proc. Ann. French Soc. Astr \& Ap.}, Reyl\'e, Richard, Cambr\'esy, Deleuil, P\'econtal, Tresse \& Vaughlin, eds.,  p.237\\
Bessell, M. 2005, ARAA, 43, 293\\
Bessell, M. \& Brett, J. 1988, PASP, 100, 1134\\
Buser, R, \& Kurucz, R. 1992, A\&A, 264, 557\\
Di Benedetto, G. 1998, A\&A, 339, 858 \\ 
Di Benedetto, G. 2005, MNRAS, 357, 174\\
Elgueta, S., Graczyk, D. \& Gieren, W. 2018, ASP Conf Series, 514, 119\\
Gustafsson, B. et al 2008, A\&A, 486, 951\\
Johnson, H. 1966, ARAA, 4, 201\\
Lester, J. \& Neilson, H. 2008, A\&A, 491, 633\\
Mould, J., Clementini, G. \& Da Costa, G. 2019, PASA, 36, 1\\
Onozato, H. et al 2019, MNRAS, 486, 5600\\
Pietrzynski, G. et al 2019, Nature 567, 200\\

\end{document}